\documentclass[11pt]{article}
\usepackage[dvips]{graphicx}
\usepackage{lscape}
\newcommand\nucl[2]{{}^{#1}{\rm #2}}
\newcommand{\geff}[1]{g_{#1}}

\begin{document}

\setcounter{page}{0}
\vspace{3cm}
\hfill
{\large {\bf PSI\,-\,PR\,-\,01\,-\,09}}\\
\hspace*{\fill}
{\large April 2001}
\vspace{5cm}
\begin{center}
{\Large \bf
 MICROSCOPIC CALCULATION OF TOTAL
 ORDINARY MUON CAPTURE RATES
 FOR MEDIUM - WEIGHT AND HEAVY NUCLEI
}\\
\vspace{1cm}
V.~A.\ Kuz'min$^{1}$,
T.~V.\ Tetereva$^{2}$ 
K.\ Junker$^{3}$ 
and A.~A.\ Ovchinnikova$^{2}$ \\
 {\it
   ${}^1$~Joint Institute for Nuclear Research, Dubna,
   Moscow region, 141980, Russia} \\ 
 {\it
 ${}^2$~Skobeltsyn Institute of Nuclear Physics,
   Lomonosov Moscow State University, Moscow, Russia} \\ 
 {\it
 ${}^3$~Paul Scherrer Institut, CH-5232 Villigen-PSI, Switzerland} \\ 
\end{center}
\vspace{5cm}
\hspace*{8cm}
{\large 
PAUL SCHERRER INSTITUT} \\
\hspace*{8cm} {\large CH - 5232 Villigen PSI} \\
\hspace*{8cm} {\large Telephon 0041 56 310 2111} \\
\hspace*{8cm} {\large Telefax  0041 56 310 2199}\\*[1cm]
\newpage
\thispagestyle{empty}

\vspace*{3cm}

\begin{center}
{\Large \bf Microscopic Calculation of Total Ordinary Muon \\Capture 
   Rates for Medium -  Weight and Heavy Nuclei}  \\ [3mm]

{ V.~A.\ Kuz'min$^{1}$, 
  T.~V.\ Tetereva$^{2}$,
  K.\ Junker$^{3}$,
  and A.~A.\ Ovchinnikova$^{2}$
}\\ [2mm]

{\small
{\it
 ${}^1$ Joint Institute for Nuclear Research, Dubna, Moscow region, 141980, 
 Russia \\
 ${}^2$ Skobeltsyn Nuclear Physics Institute,
 Lomonosov Moscow State University, \\ Moscow, Russia \\
 ${}^3$ Paul Scherrer Institute,
 CH-5232 Villigen-PSI, Switzerland \\
}}

\end{center}
\noindent
\begin{center}
{\bf Abstract}
\end{center}
{\small
\noindent
 Total Ordinary Muon Capture (OMC) rates are calculated
 on the basis of the Quasiparticle Random Phase Approximation
 for several spherical nuclei from $\nucl{90}{Zr}$ to $\nucl{208}{Pb}$.
 It is shown that total OMC rates calculated with the free value
 of the axial-vector coupling constant $\geff{A}$ agree well with
 the experimental data for medium-size nuclei and exceed
 considerably the experimental rates for heavy nuclei.
 The sensitivity of theoretical OMC rates to the nuclear
 residual interactions is discussed.}

\vfil
\newpage

\section{Introduction}

 The main aim of investigating muon capture on atomic nuclei
 is the determination of the coupling constants of the weak
 hadronic current in the nuclear environment.
 In principle one is interested in the weak axial-vector
 $\geff{A}$ and pseudoscalar $\geff{P}$ coupling constants.
 The observables of nuclear muon capture are calculated as
 functions of $\geff{A}$ and $\geff{P}$ and then compared
 to the corresponding experimental data.
 The Radiative Muon Capture (RMC) is traditionally considered
 as the most promising source of information on $\geff{P}$
 \cite{TH1995}.
 The sensitivity to $\geff{P}$ could be increased if one
 considers the ratio of the total RMC rate to the total Ordinary Muon
 Capture (OMC) rate \cite{GOT1986,EKT1998}.
 Therefore the problem arises how to make consistent calculations
 of the total RMC and OMC rates within the same nuclear model.

 The investigation of total OMC rates is also of interest in itself
 because it allows the determination of $\geff{A}$ independent
 from beta-decay.
 In contrast to beta-decay all the final nuclear states with
 noticeable transition strength can be populated in muon
 capture.
 Therefore, small variations in the low-energy parts of theoretical
 strength functions give no dramatic changes in the calculated
 OMC rates in contrast to calculations of $\log (ft)$ values.
 An additional advantage of muon capture as compared to beta decay
 is that the total OMC rates are measured for many stable and
 long-living nuclei.
 Therefore, one can in OMC not only study the variation of
 $\geff{A}$ with nuclear mass, but more delicate effects,
 such as the isotopic dependence of the effective weak
 interaction constants between leptons and nucleons.

 The present situation in the field of OMC rate measurements can be
 summarized as follows.
 Total OMC rates have been measured for many nuclei with high
 precision \cite{SMR1987}. The theoretical interpretation of the experimental 
 information is, however, still controversial.
 The main reason lies in the necessity of a correct description
 of the nuclear response in both, OMC and RMC.

 The theoretical investigation of nuclear muon capture has
 a rather long history.
 Up to now, three different approaches have been developed for
 calculating total (exclusive) OMC rates on complex nuclei.
 The first approach is based on the closure approximation and
 related sum rule methods.
 In both cases the energy of the outgoing neutrino is replaced
 by some average value which is a parameter of the theory.
 The OMC rate is then obtained using the closure relation for
 the final nuclear states \cite{GP1974,NBGM1987}.
 The second approach utilizes the local density approximation.
 Here the OMC rates are calculated for infinite and uniform
 nuclear matter as a function of the proton and neutron densities.
 The OMC rate for a finite-size nucleus is then obtained
 by integrating this function over the realistic density
 distribution or by determining its value for a certain value of the
 nuclear matter density \cite{COF1990,FW1992}.
 The common drawback of both approaches is that nuclear muon
 capture is considered without any connection to other processes
 which may occur in a nucleus.
 Also, the collective nature of the nuclear response to external
 fields is lost.
 The third approach, which is used in this paper, does not suffer
 from these defects.
 Here the (exclusive) OMC rates are calculated for all definite
 final states of the product nucleus and the total rate is obtained
 by summing over all considered final states.
 \begin{equation}
 \label{eq:tot_rates}
  \Lambda_{\rm tot} = \sum_{f} \Lambda_{fi} .
 \end{equation}
 The calculations of Refs.\ \cite{EKT1998,Bun1966,AueKl,UV1992,KLV2000}
 have been carried out within this approach.

 In the present work we study the total OMC rates on heavy nuclei within
 a microscopic description of the nuclear structure.
 The one-body effective Hamiltonian of nuclear OMC obtained
 within the Morita and Fujii formalism \cite{MorFuj} is used.
 The wave function of the bound muon and its binding
 energy are calculated approximately \cite{Pustovalov}, 
 taking into account the finite size of the nucleus.
 The nuclear matrix elements of the effective OMC Hamiltonian
 and the excitation energies of the states of a product nucleus
 are calculated within the Quasiparticle Random Phase
 Approximation (QRPA), an extension of the usual RPA
 to nonclosed shell nuclei.
 Velocity-dependent terms are included in the calculations
 and have been evaluated with Woods-Saxon single-particle wave
 functions having the correct asymptotic behaviour.

\section{The Effective Hamiltonian of Nuclear Muon Capture}

 The total rate of OMC is calculated by summing the rates
 $\Lambda_{fi}$ for all partial transitions
 $ i \rightarrow f $.
 In a spherically symmetric nucleus $\Lambda_{fi}$ is
 given by \cite{BKE}
\begin{equation}
 \label{eq:part_rate}
 \matrix{
 \Lambda_{fi} = & \displaystyle
 [G \cos{\Theta_{C}}]^2 \, (E_{\nu})^{2}
   (1 - { E_{\nu} \over {M_{i} + m_{\mu}} } ) \,
   { { 2 J_{f} + 1 } \over { 2 J_{i} + 1 } } \hfill \cr
   {} &  \displaystyle \sum_{u} \bigl (
   M^{2}_{u} (u) + M^{2}_{u} (u+1)
   + M^{2}_{u} (-u) + M^{2}_{u} ( -u - 1)
   \bigr )  \hfill \cr
  }
\end{equation}
 The $M_{u}(\kappa)$'s
 are the amplitudes for the transition in which a
 neutrino is created in a state with energy $E_{\nu}$ and
 angular quantum number $\kappa$ ($\kappa = l$ for $j = l - 1/2$
 and $ \kappa = -l -1 $ for $j = l + 1/2$). 
 $u$ is the angular momentum transferred to the nucleus.
 These amplitudes are combinations of the weak form factors with
 nuclear matrix elements \pagebreak
 \begin{equation}
 \label{eq:amplitudes}
 \matrix{
 \hspace*{\fill}M_{u}(u)\quad = & 
  \displaystyle %
   \sqrt{ 2 \over { 2 u + 1 } } \biggl (
   \sqrt{u} G_{V} [0uu]
   -\sqrt{{u+1} \over 3} G_{A} [1uu] \hfill \cr
 \noalign{\vskip 1 true pt}
   & \displaystyle %
   -\sqrt{{2u + 1} \over 3 } { g_{V} \over M } [1u-1up]
   \biggr )  \hfill \cr
 \noalign{\vskip 2 true pt}
 \hspace*{\fill}M_{u}(-u - 1)\quad = & \displaystyle %
   \sqrt{ 2 \over { 2u + 1}} \biggl (
   \sqrt{u+1} G_{V} [0uu]
  + \sqrt{ u \over 3 } G_{A} [1uu] \hfill \cr
 \noalign{\vskip 1 true pt}
   & \displaystyle %
  - \sqrt{{2u + 1} \over 3 } { g_{V} \over M } [1u+1up]
  \biggr )  \hfill \cr
 \noalign{\vskip 2 true pt}
 \hspace*{\fill}M_{u} (-u)\quad = & \displaystyle %
   \sqrt{ 2 \over {2u+1}} \biggl (
   - \sqrt{ {2u+1} \over 3 }
      (G_{A} - {u \over {2u+1}} G_{P}) [1u-1u]
  \hfill \cr
 \noalign{\vskip 1 true pt}
 {} & \displaystyle %
   + \sqrt{ {u(u+1)} \over {3 (2u+1)} } G_{P} [1u+1u]
   - \sqrt{u} { g_{A} \over M} [0uup]
  \hfill \cr
 \noalign{\vskip 1 true pt}
 {} & \displaystyle %
   + \sqrt{ {u+1} \over 3 } { g_{V} \over M } [1uup]
   \biggr )  \hfill \cr
 \noalign{\vskip 2 true pt}
 \hspace*{\fill}M_{u} (u + 1)\quad = & \displaystyle %
   \sqrt{ 2 \over {2u+1} } \biggl (
   - \sqrt{ {u(u+1)} \over {3(2u+1)}} G_{P} [1u-1u]
   + \sqrt{ {2u+1} \over 3 }
      (G_{A}
   \hfill \cr
 \noalign{\vskip 1 true pt}
   {} & \displaystyle %
    - { {u+1} \over {2u+1} } G_{P} ) [1u+1u]
   + \sqrt{ u+1 } { g_{A} \over M } [0uup]
   + \sqrt{ u \over 3} { g_{V} \over M } [1uup]
   \biggr ) \hfill \cr
 }
 \end{equation}
 Here the effective form factors are defined in the usual way
\begin{equation}
 \matrix{
\hspace*{\fill}G_{V}\quad = &  g_{V}(q^2) \bigl ( 1 + E_{\nu} / 2M \bigr )
     + g_{S}(q^2) \hfill \cr
\hspace*{\fill}G_{A}\quad = &  g_{A}(q^2) - \bigl ( g_{V}(q^2) + g_{M}(q^2) \bigr )
           \bigl ( E_{\nu} / 2 M \bigr ) \hfill \cr
\hspace*{\fill}G_{P}\quad = &  \bigl ( g_{P}(q^2) - g_{A}(q^2) - g_{T}(q^2)
            - g_{V}(q^2) - g_{M}(q^2) \bigr )
            \bigl (E_{\nu} / 2 M \bigr )  \hfill \cr
 }
\end{equation}
 and the nuclear single-particle matrix elements are given by
 \begin{equation}
 \label{eq:brackets}
 \matrix{
\hspace*{\fill}[0uu]\quad = & %
 \displaystyle
   \bigl \langle J_{f} \parallel \sqrt{ 1 \over {4 \pi } }
    \sum_{k = 1 }^{A} \phi_{\mu}(r_{k}) \,
       t^{+}_{k} \, j_{u}(E_{\nu} r_{k}) \, Y_{u}(\hat{r}_{k})
       \parallel J_{i} \bigr \rangle / \sqrt{ 2 J_{f} + 1 } \hfill \cr
 \noalign{\vskip 1 true pt}
\hspace*{\fill}[1wu]\quad = & %
 \displaystyle
   \bigl \langle J_{f} \parallel \sqrt{ 3 \over { 4 \pi } }
    \sum_{k = 1}^{A} \phi_{\mu}(r_{k}) \, t^{+}_{k} \, j_{w}(E_{\nu} r_{k} )
     \, [\sigma \otimes Y_{w}(\hat{r}_{k}) ]_{u}
      \parallel J_{i} \bigr \rangle / \sqrt{ 2 J_{f} + 1 } \hfill \cr
 \noalign{\vskip 1 true pt}
\hspace*{\fill}[1wup]\quad = & %
  \displaystyle
    i \bigl \langle J_{f} \parallel \sqrt{ 3 \over { 4 \pi } }
     \sum_{k=1}^{A} \phi_{\mu}(r_{k}) \, t^{+}_{k} \,
      j_{w}(E_{\nu} r_{k}) \, [ Y_{w}(\hat{r}_{k}) \otimes p_{k} ]_{u}
      \parallel J_{i} \bigr \rangle / \sqrt{ 2 J_{f} + 1 } \hfill \cr
 \noalign{\vskip 1 true pt}
\hspace*{\fill}[0uup]\quad = & %
   \displaystyle
    i \bigl \langle J_{f} \parallel \sqrt{ 1 \over { 4 \pi } }
     \sum_{k=1}^{A} \phi_{\mu}(r_{k}) \, t^{+}_{k} \,
      j_{u}(E_{\nu} r_{k}) \, Y_{u}(\hat{r}_{k}) \,
        (\vec{p}_{k}, \vec{\sigma}_{k} )
      \parallel J_{i} \bigr \rangle / \sqrt{ 2 J_{f} + 1 }  \hfill \cr
 }
 \end{equation}
 where
 $ j_{u}(x) = \sqrt{ \pi / 2x } J_{u+1/2}(x) $ 
 is a spherical Bessel function and
\begin{equation}
 [\sigma \otimes Y_{w}(\hat{r}) ]_{u,m_{u}} =
  \sum_{m, m_{w}} \langle 1 m \, w m_{w} \vert u m_{u} \rangle \,
  \sigma_{m} Y_{w, m_{w}}(\hat{r})
\end{equation}
 is the tensor product of two spherical tensor operators. 
 $\phi_{\mu}(r)$ is the radial wave function of the bound muon.
 For the nucleon isospin operators $t^{+}$ the convention 
 $ t^{+} \vert p \rangle = \vert n \rangle $ is used.

 Eqs.(\ref{eq:part_rate}) and (\ref{eq:brackets}) show that the capture rate 
 depends rather strongly on the energy of the outgoing neutrino
\begin{equation}
  E_{\nu} = ( m_{\mu} - \vert \epsilon_{1S} \vert + M_{i}
            - M_{f} - E^{\ast} )
 \bigl ( 1 - { { m_{\mu} - \vert \epsilon_{1S} \vert + M_{i}
        - M_{f} - E^{\ast} } \over { 2 ( m_{\mu} + M_{i} ) } } \bigr ) 
\end{equation}
 For a large $Z$ nucleus the muon binding energy $\epsilon_{1S}$
 has to be calculated taking into account the finite size of the
 nuclear charge distribution.
 The excitation energy of the final nuclear state $E^{\ast}$
 enters into the capture rate mainly through the neutrino energy.
 In order to obtain the transition energies and nuclear matrix
 elements (\ref{eq:brackets}) one has to use some nuclear model.
 In the present work the QRPA has been used since it gives a
 precise prescription of how the interaction between particle
 and hole excitation can be included in the calculation of the
 nuclear matrix elements (\ref{eq:brackets}) and the excitation spectrum of
 the product nucleus.

\section{The Nuclear Model}

 A detailed description of the nuclear model used in the present work can be
 found in \cite{EKT1998} and will not be repeated here. The nuclear 
 Hamiltonian consists of a mean field part, a monopole pairing interaction
 between like particles and a residual interaction. 
\begin{equation}
  H_{M} = \sum_{t_{3} = \pm 1/2} \biggl (
         H_{\rm mean}(t_{3}) + H_{\rm pair}(t_{3}) \biggr )
        + H_{\rm resid} 
\end{equation}
 For protons and neutrons separate Woods-Saxon potentials including spin-orbit
 interactions have been used to approximate the mean field. The mean field
 and pairing Hamiltonians
 \begin{equation}
 \label{eq:mean}
 \matrix{ \displaystyle
  H_{0}( t_{3} ) = & \displaystyle
  H_{\rm mean}(t_{3}) + H_{\rm pair}(t_{3}) \hfill \cr
\noalign{\vskip 1 true pt}
 \hfill = & \displaystyle
  \sum_{j, m} 
   E_{j t_{3}} a^{\dag}_{j m t_{3}} a_{j m t_{3} }
   - { G_{t_{3}} \over 4 }
   \sum_{j m, \, j^{\prime} m^{\prime}}
   (-1)^{j - m + j^{\prime} - m^{\prime} }
   a^{\dag}_{j m t_{3}}
   a^{\dag}_{j, -m t_{3}}
   a_{j^{\prime}, -m^{\prime} t_{3}}
   a_{j^{\prime}, m^{\prime} t_{3}} \hfill \cr
 }
 \end{equation}
 are in a first step approximately diagonalized by the special Bogoliubov
 transformation (see e.g. \cite{Solov1976})
\begin{equation}
\label{eq:bogol}
  a_{j m t_{3}} =
    u_{j t_{3}} \alpha_{j m t_{3}} +
    (-1)^{j - m}
    v_{j t_{3}} \alpha^{\dag}_{j, -m t_{3}} .
\end{equation}
 Solving the BCS equations leads to the Independent Quasiparticle Hamiltonian 
\begin{equation}
  H_0(t_{3}) \rightarrow \sum_{j m}
            \epsilon_{j t_{3}} \alpha^{\dag}_{j m t_{3}}
                            \alpha_{j m t_{3}} \,
\end{equation}
 with
\begin{equation}
   \epsilon_{j t_{3}} = \sqrt{ ( E_{j t_{3}} -\lambda_{t_{3}} )^2 +
                    C^2_{t_{3}} }
 \quad \hbox{\rm and} \quad
  C_{t_{3}} = { G_{t_{3}} \over 2 }
        \sum_{j, m} \, u_{j t_{3}} v_{j t_{3}} \, .
\end{equation}
 An averaging over the quasiparticle vacuum state 
 $ \alpha_{j m t_{3} } \vert 0 \rangle = 0 $ is implied.
 As residual interaction we use an effective
 isospin-invariant separable interaction of the form
 \begin{equation}
 \label{eq:residual}
 \matrix{
  H_{\rm resid} = & \displaystyle
  - { 1 \over 2 } \sum_{L,M}
    ( \kappa^{L}_0 +\kappa^{L}_1 (\vec{\tau}_1 \cdot \vec{\tau}_2 ) )
             Q^{\dag}_{LM}(1) Q_{LM}(2)  \hfill \cr
 \noalign{\vskip 1 true pt}
  & \displaystyle
  - { 1 \over 2 }  \sum_{L,J,M} ( \kappa^{LJ}_0 + \kappa^{LJ}_1
            (\vec{\tau}_1 \cdot \vec{\tau}_2 ) )
            Q^{\dag}_{LJM}(1) Q_{LJM}(2) \ . \hfill \cr
 }
 \end{equation}
 where $Q_{LM}$ and $Q_{LJM}$ are single-particle multipole and spin-multipole
 operators and the isospin structure is displayed explicitly. Using 
\begin{equation}
\label{eq:pauli}
  ( \vec{\tau}_1 \cdot \vec{\tau}_2 ) = 4 ( \vec t_{1} \cdot \vec t_{2} ) = 
4 t^0_{1} t^0_{2}  + 2 (t^{-}_1 t^{+}_2 + t^{+}_1 t^{-}_2).
\end{equation}
one can combine the isospin operators with the multipole and spin-multipole 
operators leading to a new set of single-particle operators 
\begin{equation}
\label{eq:multip}
  Q^{\rho}_{LM}(k) = \sum_{j^{\prime} m^{\prime} t^{\prime}_3, j m t_3}
          \langle j^{\prime} m^{\prime} t^{\prime}_3 \vert
          i^{L} f_{L}(r_{k}) \, Y_{LM}(k) \, t^{\rho}_k \vert j m t_3 \rangle
           a^{\dag}_{j^{\prime} m^{\prime} t^{\prime}_3}
           a_{j m t_3}
\end{equation}
 and
\begin{equation}
\label{eq:spin_multip}
    Q^{\rho}_{LJM}(k) = \sum_{j^{\prime} m{\prime} t^{\prime}_3, j m t_3}
      \langle j^{\prime} m^{\prime} t^{\prime}_3 \vert
        i^{L}  f_{LJ}(r_{k}) \, [ Y_{L}(k) \otimes \sigma(k) ]_{JM} \, 
        t^{\rho}_k  \vert j m t_3 \rangle
           a^{\dag}_{j^{\prime} m^{\prime} t^{\prime}_3}
           a_{j m t_3} \ 
\end{equation}
 where
 $t_k^\rho$ is an isospin operator out of the set \{$\hat{1}_k$, 
 $t_k^{0}=t_k^{z}$, $t_k^{\pm}=t_k^{x}\pm i t_k^{y}$, ($k=1,2$)\}. 
 In the literature \cite{VdoSol83} one finds two most frequently used variants
 of the radial form factors $f_{L}(r)$ and $f_{LJ}(r)$. These are
 \begin{equation}
 \label{eq:rL}
    f_{L}(r) = f(r)_{LJ} = r^{L}.
 \end{equation}
 and 
 \begin{equation}
 \label{eq:dudr}
   f_{L}(r) = f(r)_{LJ} = f(r) = {d \over dr} U(r) ,
 \end{equation}
 where $U(r)$ is the central part of the shell-model potential used in 
 $H_{mean}$. The mixed products $t_1^{-} t_2^{+}$ and $t_1^{+} t_2^{-}$ in
 Eq.(\ref{eq:pauli}) lead to particle-hole excitations changing the charge
 of the nucleus and are therefore involved in a description of charge-exchange 
 interaction processes such as $\beta$-decay, $\mu$-capture or direct 
 $(p,n)$ and $(n,p)$ reactions. The corresponding parts in the 
 Hamiltonian (\ref{eq:residual}) are
 constructed with the singl-particle operators
\begin{equation}
  \Omega_{JM} = \sum_{j_{n}m_{n}, \, j_{p}m_{p}}
       \langle j_{n} m_{n} \vert O_{JM} t^{+} \vert j_{p} m_{p} \rangle
      \, a^{\dag}_{j_n,m_n} \, a_{j_p,m_p}
\end{equation}
 and their Hermitian conjugates, where 
 $O_{JM}$ can be either $ i^{J} f_{J}(r)Y_{JM}(\hat{r}) $ or \newline
\mbox{ $ i^{L} f_{LJ}(r) [Y_L(\hat{r}) \otimes \sigma ]_{JM} $} respectively.

 The residual interaction (\ref{eq:residual}) contains only scalar products
 of the form \break
 \hbox{ $ \bigl ( [Y_{J-1}(\hat{r}_{1}) \otimes \sigma_{1}]_{J}, \,
  [Y_{J-1}(\hat{r}_{2}) \otimes \sigma_{2}]_{J} \bigr ) $ }
 and
 $ \bigl ( [Y_{J+1}(\hat{r}_{1}) \otimes \sigma_{1}]_{J}, \,
  [Y_{J+1}(\hat{r}_{2}) \otimes \sigma_{2}]_{J} \bigr ) $.
 The tensor interaction which would mix the 
  $ [Y_{J-1}(\hat{r}_{1}) \otimes \sigma_{1}]_{JM} $  and
  $ [Y_{J+1}(\hat{r}_{2}) \otimes \sigma_{2}]_{JM} $ terms 
 has been omitted because its inclusion would slightly influence 
 the properties of the charge-exchange resonances in 
 the range where the coupling constants assume reasonable values. 

 To achieve an approximate diagonalization of the residual interaction
 Hamiltonian in the QRPA one introduces phonon creation and
 destruction operators. 
 They are defined as linear combinations of tensor products of the 
 quasiparticle operators $\alpha_{j m t_{3}}$ and 
 $\alpha_{j^{\prime} m^{\prime} t_{3}^{\prime}}$ and their Hermitian 
 conjugates (Eq.\ref{eq:bogol}). 
\begin{equation}
\label{eq:phonon}
  \Omega^{i}_{JM} =
  \sum_{j_p, j_n} \biggl (
    \psi^{i}_{j_p, j_n} [\alpha_{j_p} \otimes \alpha_{j_n}]_{J,M}
   -(-1)^{J-M} \phi^{i}_{j_p, j_n}
     [\alpha_{j_p} \otimes \alpha_{j_n}]^{\dag}_{J,-M} \biggr )  
\end{equation}
 The phonon amplitudes 
 $\psi^{i}_{j_n, j_p}$ and $\phi^{i}_{j_n, j_p}$ are orthonormalized 
 according to
\begin{equation}
  \Phi(i,i^{\prime}) \equiv
  \sum_{j_p, j_n}\{ \psi^{i}_{j_p, j_n} \psi^{i^{\prime}}_{j_p, j_n}
           - \phi^{i}_{j_p, j_n} \phi^{i^{\prime}}_{j_p,j_n} \} =
       \delta_{i,i^{\prime}}. 
\end{equation}
 Having expressed the residual interaction in terms of the phonon operators
 (\ref{eq:phonon}) one obtains the QRPA equations through a 
 variational principle \cite{Solov1976}
\begin{equation}
\label{eq:var}
 \delta \biggl \{ \langle \vert \Omega^{i}_{JM} H_M \,
       {\Omega^{i}_{JM}}^{\dag} \vert \rangle
        - \langle \vert H_M \vert \rangle
           - \omega_{i} \, (\Phi(i,i) - 1) \biggr \} = 0  
\end{equation}
 using the normalization of the phonon amplitudes as a subsidiary condition.
 $\vert \rangle$ is the phonon vacuum:
 $ \Omega^{i}_{JM} \vert \rangle = 0   $, approximating the ground state of a
 double even nucleus. The QRPA equations resulting from (\ref{eq:var}) are a
 system of homogeneous linear equations determining the phonon amplitudes
 $\psi^{i}_{j_n, j_p}$ and $\phi^{i}_{j_n, j_p}$ and the excitation energies 
 $\omega_i$. 
\begin{equation}
 \matrix{
  \displaystyle
  R^{+}_{q,q^{\prime}} g^{i}_{q^{\prime}} & %
    - & \omega_{i} w^{i}_{q} & = 0 , \cr
  \noalign{\smallskip}
  \displaystyle
 -\omega_{i} g^{i}_{q}  & %
 + & R^{-}_{q,q^{\prime}} w^{i}_{q^{\prime}} & = 0 , \cr
 }
\end{equation}
 where the following abbreviations have been used
\begin{equation}
  g^{i}_{q} = \psi^{i}_{j_p, j_n} + \phi^{i}_{j_p, j_n} ,
  \qquad
  w^{i}_{q} = \psi^{i}_{j_p, j_n} - \phi^{i}_{j_p, j_n} ,
  \qquad
  q \equiv (j_{p}, j_{n}) ,
\end{equation}
\begin{equation}
   \epsilon_{q} = \epsilon_{j_p} + \epsilon_{j_n} ,
   \qquad
   u^{\pm}_{q} = u_{j_p} v_{j_n} \pm v_{j_p} u_{j_n}
\end{equation}
\begin{equation}
  R^{\pm}_{q, q^{\prime}} = \epsilon_{q} \delta_{q,q^{\prime}}
  - { 2 \over 2 J + 1 } \biggl ( \kappa^{J}_{1}
   h^{J}_{q} u^{\pm}_{q}  \, h^{J}_{q^{\prime}} u^{\pm}_{q^{\prime}}
  +  \kappa^{JJ}_{1}
   h^{JJ}_{q} u^{\pm}_{q} \, h^{JJ}_{q^{\prime}} u^{\pm}_{q^{\prime}}
   \biggr )
\end{equation}
 for naturaly parity states and
\begin{equation}
  R^{\pm}_{q, q^{\prime}} = \epsilon_{q} \delta_{q,q^{\prime}}
  - { 2 \over 2 J + 1 }
   \biggl ( \kappa^{J-1,J}_{1}
   h^{J-1,J}_{q} u^{\pm}_{q} \,
   h^{J-1,J}_{q^{\prime}} u^{\pm}_{q^{\prime}}
   + \kappa^{J+1,J}_{1}
   h^{J+1,J}_{q} u^{\pm}_{q} \,
   h^{J+1,J}_{q^{\prime}} u^{\pm}_{q^{\prime}}
   \biggr )
\end{equation}
 for unnatural parity states. $h^J_{q}$ and $h^{LJ}_{q}$ stand for the
 reduced matrix elements of the multipole (\ref{eq:multip}) and spin-multipole 
 (\ref{eq:spin_multip}) single-particle operators with $t^{\rho} = t^{-}$. 
 The QRPA amplitudes for the transitions from the even-even ground state
 to the excited states with total spin $JM$,
 and energies $\omega_i \pm (\lambda_n - \lambda_p)$ are given by
\begin{equation}
  b^{+}_{JM}(i) = { 1 \over \sqrt{ 2 J + 1 } }
  \sum_{j_p, j_n} \langle j_p \Vert O_J t^{-} \Vert j_n \rangle
  ( v_{j_p} u_{j_n} \psi^{i}_{j_p, j_n} +
       u_{j_p} v_{j_n} \phi^{i}_{j_p, j_n} )
\end{equation}
 if the charge of the nucleus decreases from $Z$ to $Z-1$
 (as, for example, in $(n,p)$ reactions) or by
\begin{equation}
  b^{-}_{JM}(i) = { 1 \over \sqrt{ 2 J + 1 } }
  \sum_{j_p, j_n} \langle j_p \Vert O_J t^{-} \Vert j_n \rangle
  ( u_{j_p} v_{j_n} \psi^{i}_{j_p, j_n} +
       v_{j_p} u_{j_n} \phi^{i}_{j_p, j_n} )
\end{equation}
 if the charge increases from $Z$ to $Z+1$ during the transition
 (as in $(p,n)$ reactions).

 In the following we will make a few remarks on the nuclear model parameters. 
 There are three kinds of parameters which characterize the nuclear structure 
 of our model Hamiltonian. The parameters of the mean field 
 (Woods-Saxon potential) are determined such that they reproduce best the
 single-particle excitations of neighbouring odd-mass nuclei. Different 
 potentials have been used for protons and neutrons. The parameters of the
 monopole pairing interaction have been obtained from the odd-even effect of 
 nuclear binding energies. Since the theoretical binding energies depend on
 the single-particle energies there is a certain correlation between the
 pairing constants and the parameters of the Woods-Saxon potential. Usually
 one set of monopole pairing constants and Woods-Saxon potential parameters
 has been used in the calculations for a whole group of neighbouring nuclei.
 These parameter sets have been taken from \cite{Pon79}. The residual 
 interaction between particle-hole excitations causes the collective small 
 amplitude vibrations. The effective constants of this interaction have been 
 determined by comparing the calculated excitation energies and the transition 
 strengths of the collective isovector states with experiment. In the present
 work only the isovector interaction is of interest since we are dealing 
 with a charge changing process. A detailed discussion of the used form of
 separable residual interaction can be found in the work of \cite{VdoSol83}. 
 In order to demonstrate the sensitivity of the calculated total muon capture
 rate ${\Lambda}_{tot}$ to the residual interaction, both variants of radial 
 form factors have been used with several values for the isovector coupling 
 constant.
\newpage

\section{Results of the Calculations}

 In this section we present detailed results of the calculated total OMC rates
 obtained for spherical nuclei of different mass regions. The data are 
 presented in Table \ref{tab:zr90gt} to Table \ref{tab:total_rates}. The 
 capture rates shown in \hbox{Fig.\ \ref{fig:zr90omc}} and \hbox{Fig.\ 
 \ref{fig:pb208omc}} are presented in the form of running sums 
 \begin{equation}
 \label{eq:run_rate}
   \Lambda (E) = \sum_{k: \, E_{k} < E}  \Lambda_{k}, 
 \end{equation}
 where the energies are measured with respect to the ground state of the 
 initial nucleus. In 
 order to get a feeling for the sensitivity of the theoretical muon
 capture rates on the chosen shape and strength of the residual interaction, 
 all calculations have been performed for both types of interaction 
 (\ref{eq:rL}) and (\ref{eq:dudr}). Throughout we have employed 
 $g_{P}/g_{A} = 6.0$ in the calculations.

\subsection{$\nucl{\bf 90}{\bf Zr}$ and $\nucl{\bf 92}{\bf Mo}$}

 The distribution of the strength of the transition operator $\sigma t^{-}$ 
 over the excitation energies (Gamow-Teller strength function) 
 has been studied in detail for the reactions \break
 $ \nucl{90}{Zr}(p,n)\nucl{90}{Nb} $ \cite{zr90pn,zr90high}
 and $\nucl{90}{Zr}(\nucl{6}{Li},\nucl{6}{He})\nucl{90}{Nb}$
 \cite{zr90ion}.
 The strength function has a prominent peak at an energy of $15.6$ MeV.
 For this peak the reduced probability of the GT transition $B(GT) = 10$.  
 The total observed transition strength below $20$ MeV excitation energy
 is around $20$.
 The total GT strength should be larger than $3 ( N - Z ) = 30$,
 the value given by the Ikeda sum rule \cite{Ikeda63}.

 Table \ref{tab:zr90gt} shows how the calculated properties
 of the GT strength function depend on $\kappa^{01}_{1}$ and
 $\kappa^{21}_{1}$.
 The model of non-interacting quasiparticles (residual interaction
 switched off: $ \kappa^{01}_{1} = \kappa^{21}_{1} = 0 $)
 cannot give the correct position of the peak of the strength function.
 For $L=0$ and $J=1$, the residual interaction (\ref{eq:rL})
 reduces to the simple $(\vec{\sigma}, \vec{\sigma})$ interaction
 considered in \cite{Gaarde1981}.
 The calculations with
 $\kappa^{01}_{1} = -23/A$ recommended in \cite{Gaarde1981} correctly
 reproduce the position of the maximum of the strength function.
 The results of calculations with slightly different constants
 $\kappa^{01}_{1} = -25/A$ and $\kappa^{01}_{1} = -28/A$), 
 presented in \hbox{Table \ref{tab:zr90gt}}, show that the position of
 the maximum of the strength function is sensitive to $\kappa^{01}_{1}$.
 The strength in the peak region does not depend on $\kappa^{01}_{1}$
 and considerably exceeds the experimentally observed one.
 So, the best value of the effective constant for the residual interaction
 (\ref{eq:rL}) is $\kappa^{01}_{1} = -23/A$.

 Table \ref{tab:zr90gt} shows that the calculated energy of the GT
 resonance is less sensitive to $\kappa^{01}_{1}$ for the residual
 interaction with form factor (\ref{eq:dudr}).
 The correct position of the resonance is reproduced with
 $\kappa^{01}_{1} = - 0.33/A$.
 Simultaneously the strength below and in the resonance region 
 is considerably smaller than the corresponding
 strength calculated with interaction (\ref{eq:rL})
 ($24.6$ to be compared to$29.2$) and is closer
 to the experimental value.
 The rest of the transition strength is located at high-excited
 $1^{+}$ states \cite{JKT1999}.
 Recently, new experimental data of the GT transition
 strength at very high excitation energies became available
 \cite{zr90high}.
 The total $B(GT)$ strength calculated with the residual interaction
 (\ref{eq:dudr}) is equal to $32.0$. 
 This agrees with the experimental value $34.2 \pm 1.6$
 obtained in \cite{zr90high} by a multipole decomposition of the experimental
 $\nucl{90}{Zr}(p,n)\nucl{90}{Nb}$ cross sections. 
 From this consideration one concludes that the
 residual interaction (\ref{eq:dudr}) provides a better
 description of the $\sigma t^{-}$ strength function than the
 interaction (\ref{eq:rL}).

 Recent measurements of the $\sigma t^{+}$ transition
 strength in $\nucl{90}{Zr}(n,p)\nucl{90}{Y}$ \cite{zr90np}
 can also be compared with our calculations, since in this reaction 
 the charge of the nucleus decreases as in muon capture.
 The $\sigma t^{+}$ strength, summed over all experimentally
 observed states is $B^{+}_{\Sigma}(GT) = 1.0 \pm 0.3$ \cite{zr90np}.
 Our calculation with the interaction (\ref{eq:rL}) gives
 the following distribution of the transition strength.
 A considerable part of $B^{+}_{\Sigma}(GT)$ is concentrated
 on the first $1^{+}$ state of $\nucl{90}{Y}$.
 The other $1^{+}$ states have excitation energies between
 $10$ and $15$ MeV.
 For each of these states, $B^{+}(GT) < 0.2$.

 The $\sigma t^{+}$ strength function calculated with the 
 interaction (\ref{eq:dudr}) differs from
 the strength function obtained with interaction (\ref{eq:rL}).
 The $B^{+}_{\Sigma}(GT)$ calculated with interaction
 (\ref{eq:dudr}) is almost three times the value of $B^{+}_{\Sigma}(GT)$
 obtained with interaction (\ref{eq:rL}).
 This is due to highly excited states which are absent in the 
 calculations with the $f(r) = r^{L}$ interaction.
 As before, the strongest transition goes to the first $1^{+}$ state 
 of $\nucl{90}{Y}$, but the strengths of transitions going to
 $1^{+}$ states with energies between $5$ and $15$ MeV are
 comparable to $B^{+}_{1}(GT)$.
 In this case the transition strength is distributed more uniformly over 
 the excitation energies and the shape of the theoretical strength
 function is closer to the experimental one.

 It should be noted that the energy of the first $1^{+}$ state
 calculated for each of the two residual interactions does not
 depend on $\kappa^{01}_{1}$, and the corresponding $B^{+}_{1}(GT)$
 decreases slightly whith growing $ \vert \kappa^{01}_{1} \vert $. 
 This indicates that already in $\nucl{90}{Zr}$ the neutron excess
 prevents the creation of low-lying collective $1^{+}$ states
 in $\nucl{90}{Y}$.
 The residual interaction with $f(r) = dU / dr$ (\ref{eq:dudr}) couples
 single-particle states with wave functions having the same
 orbital quantum numbers and different number of nodes in the
 radial parts.
 Due to the residual interaction (\ref{eq:dudr}), these particle-hole
 excitations interact with particle-hole states which are all members of 
 one spin-orbital multiplet and create in this way 
 high-excited collective states \cite{JKT1999}.
 The transitions to those states increase $B^{+}_{\Sigma}(GT)$.

 It was shown in \cite{EKT1998} that the theoretical 
 OMC rates are rather insensitive to the constants of the multipole
 residual interaction $\kappa^{J}_{1}$. 
 Nevertheless, the constant of the isovector monopole residual 
 interaction, $\kappa^{0}_{1}$, can be determined from analyzing 
 isobar analog states (IAS). 
 Results for the $0^{+}$ charge-exchange excitation in $\nucl{90}{Zr}$ 
 are presented in \hbox{Table \ref{tab:zr90ia}}.
 The independent quasiparticle model describes the $0^{+}$ charge-exchange
 states as a set of non-interacting two-quasiparticles states.
 The states carrying the main transition strength are
 $(0g_{9/2})_{p} (0g_{9/2})_{n}$ and
 $(1d_{1/2})_{p} (1d_{1/2})_{n}$ for the $t^{-}$ or $(p,n)$ branch and
 $(0f_{7/2})_{p} (0f_{7/2})_{n}$ and
 $(0d_{5/2})_{p} (1d_{5/2})_{n}$ for the $t^{+}$ or $(n,p)$ transitions.
 Both interactions produce a collective state in the $(p,n)$ excitation
 branch whose energy coincides with the experimental IAS energy of 
 $12.0$ MeV \cite{IASexp} ($\kappa^{0}_{1} = -0.43/A$ for interaction 
 (\ref{eq:dudr}) and $\kappa^{0}_{1} = -35.0/A$ for interaction 
 (\ref{eq:rL})). In both cases this state consumes almost all of 
 the $t^{-}$ transition strength. 

\begin{figure}
\begin{center}
\includegraphics[width=0.85\linewidth]{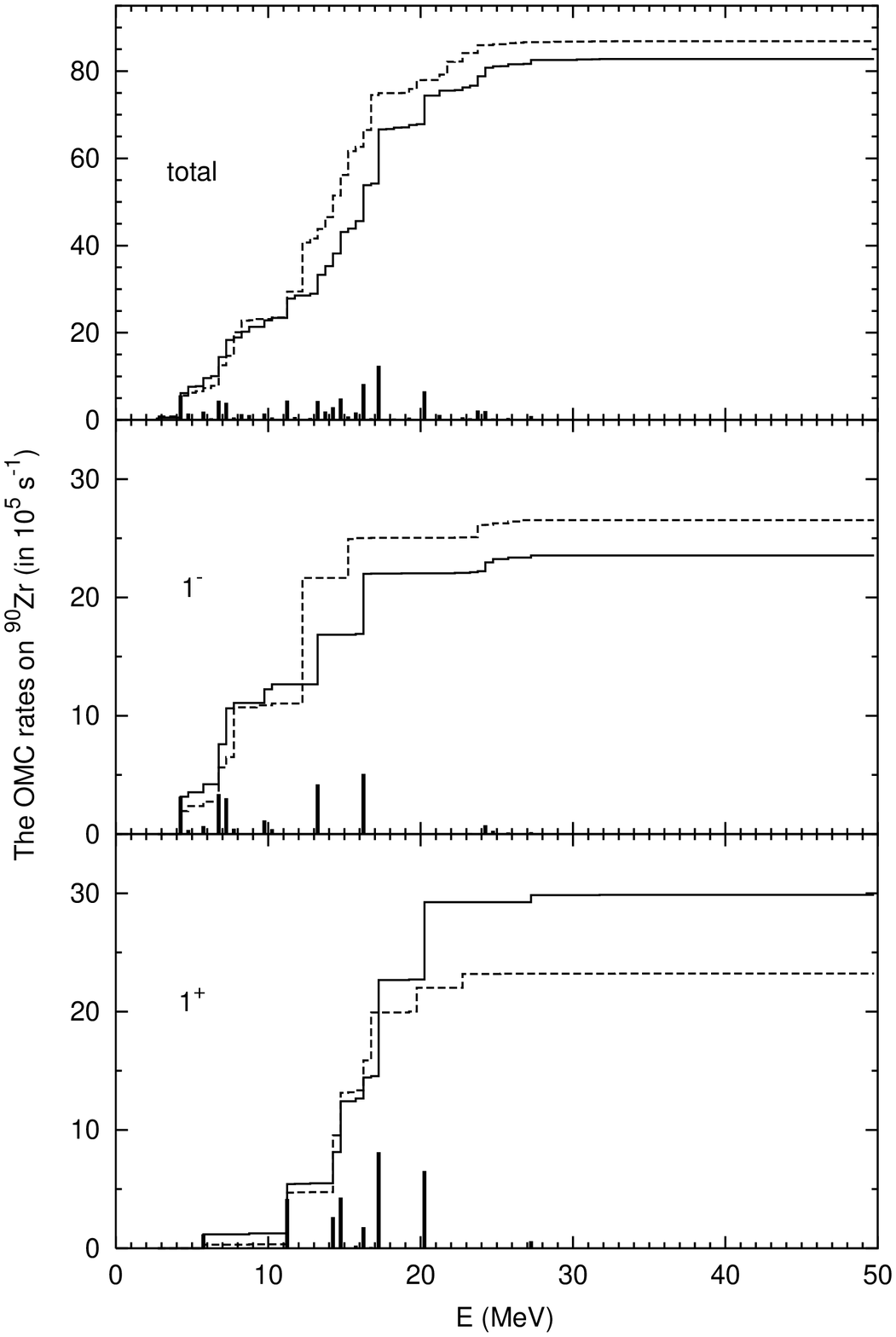}
\end{center}
\caption{ $\nucl{~90}{Zr}$ : Total and partial 
integrated capture rates up to the excitation 
energy E of the final nucleus. 
The partial rates are shown for the final states with $J^{\pi} = 1^{\pm}$.
Solid lines: results of the calculation with the residual 
interaction (\protect{\ref{eq:dudr}}),
dashed lines: the same with the interaction (\protect{\ref{eq:rL}}), the solid 
vertical bars show the distribution of the calulated partial rates  
over the excited states with $J^{\pi} = 1^{\pm}$. In the upper part of the 
figure the distribution over all excited states is shown.}
\label{fig:zr90omc}
\end{figure}

 In the $(n,p)$ branch the interaction (\ref{eq:rL}) with the above mentioned
 coupling constant is not capable to produce a collective state and the
 strength goes into the $(0f_{7/2})_{p} (0f_{7/2})_{n}$ and 
 $(0d_{5/2})_{p} (1d_{5/2})_{n}$ two-quasiparticle states. 
 The interaction (\ref{eq:dudr}) with the above mentioned coupling 
 constant is strong enough to produce a  collective state at 
 approximately the energy of the two-quasiparticle states.  
 The difference between the total $t^{-}$ and $t^{+}$ transition strengths
 is constant and does not depend on the residual interaction. 
 Therefore one can use the strength functions to determine the isovector 
 residual interaction coupling constants $\kappa^{01}_{1}$
 and $\kappa^{0}_{1}$. 
 For $L > 0$ we have used the relations
 \begin{equation}
 \label{eq:const_rL}
   \kappa^{LJ}_{1} =  { \kappa^{01}_{1}  \over
    {  \langle r^{2L} \rangle }} \qquad \mbox{for the interaction (\ref{eq:rL})}
 \end{equation}
 and 
 \begin{equation}
 \label{eq:const_dU}
  \kappa^{LJ}_{1} = \kappa^{01}_{1} \quad \qquad
  \mbox{for the interaction (\ref{eq:dudr})}
 \end{equation}
 In our theoretical total OMC rates final states with 
  $J^{\pi} = 0^{\pm}$, $1^{\pm}$, $2^{\pm}$ and $3^{\pm}$ 
 have been taken into account. 
 Contributions from final states with $J > 3$ turned out 
 to be less than 1\%. In \hbox{Table \ref{tab:zr90rates}} and 
 \hbox{Fig.\ \ref{fig:zr90omc}} we present the calculated total OMC rates
 for $\nucl{90}{Zr}$. The rates shown in \hbox{Fig.\ \ref{fig:zr90omc}} were 
 calculated with both types of interaction, using $\kappa^{0}_{1} =-0.43/A$ 
 and $\kappa^{01}_{1} =-0.33/A$ for type (\ref{eq:dudr}) and 
 $\kappa^{0}_{1} =-35/A$ and $\kappa^{01}_{1} =-23/A$ for type (\ref{eq:rL}).
 It can be seen that the main difference between the rates for the two 
 different residual interactions originates from muon capture populating 
 highly excited states. This difference is biggest for the $1^{+}$ final 
 states. Both calculated total rates agree with each other to within less 
 than 5\%. However, larger differences appear in the partial rates feeding 
 states with a specific $J^{\pi}$. In general, one observes that for 
 $A \approx 90$ nuclei the total capture rates calculated with the interaction 
 (\ref{eq:rL}) is larger than that calculated with the interaction 
 (\ref{eq:dudr}).

 The experimental total capture rate of $\Lambda = 86.6.10^{5} s^{-1}$ 
 \cite{SMR1987} is in rather good agreement with both theoretical values. 
 However, one should keep in mind that the experiment has been done with 
 the natural isotope composition of $\nucl{90}{Zr}$, whereas the calulations 
 refer to a specific isotope. 

 The theoretical capture rates $\Lambda_{\rm tot}$ show only a very slight 
 sensitivity to a variation of $g_{p}/g_{a}$. The variation in 
 $\Lambda_{\rm tot}$ for $\nucl{90}{Zr}$ amounts to $85.5.10^{5} s^{-1}$ to 
 $76.8.10^{5} s^{-1}$ if $g_{p}/g_{A}$ is increased from $4.0$ to $12.0$. The 
 contribution of the velocity dependent matrix elements $[1wup]$ and $[0uup]$ 
 to $\Lambda_{\rm tot}$ is rather small. The change in $\Lambda_{\rm tot}$ is 
 less than 2\% when they are switched off. The estimated contribution of 
 (12 - 15)\% made in \cite{UV1992} could thus not be confirmed by the 
 direct calculation. 

 The interesting observation was made that the $1^{+}$ state around $5$ MeV 
 which gives a prominent contribution in the $\nucl{90}{Zr}$(n,p)$\nucl{90}
 {Y}$ reaction, contributes only very little to the total muon capture rate. 
 This is pointing to the fact that one cannot directly use matrix elements 
 from (n,p) or (p,n) reactions to analyze muon capture data. A large B(GT) 
 does not always imply a large capture rate $\Lambda_{\rm fi}$. Here the 
 explanation is that the radial matrix elements with spherical Bessel function
 $j_{0}(k_{\nu}r)$, dominating in allowed $0^{+} \rightarrow 1^{+}$ partial 
 transitions, are suppressed by the centrifugal barrier of the two-quasi 
 particle state $(0g_{7/2})_{n} (0g_{9/2})_{p}$. This different appearance 
 of the $1^{+}$ state of the product nuleus in various reactions finds its 
 explanation only within a microscopic nuclear model.

 Table \ref{tab:mo92gt} and Table \ref{tab:mo92rates} contain the results of 
 our calculation for $\nucl{92}{Mo}$. There is a satisfactory agreement between 
 the experimental data and the rates obtained with our microscopic model. 
 As in the case of $\nucl{90}{Zr}$ the two different residual interactions 
 lead to a different population of the specific $J^{\pi}$ states which after
 summing over all final states is smoothed out. A comparison with the 
 measured capture rate on natural $\nucl{}{Mo}$ seems reasonable, since 
 $\nucl{92}{Mo}$ is the lightest even $\nucl{}{Mo}$-isotope, and the capture 
 rate decreases with increasing neutron excess.

 \subsection{OMC on Even Tin Isotopes}

 In this section we consider the even tin isotopes $\nucl{116-124}{Sn}$, 
 a long chain of stable spheri\-cal nuclei. 
 Tin isotopes have a completely filled $0g_{9/2}$
 proton subshell and gradually filled $0g_{7/2}$,
 $1d_{5/2}$, $1d_{3/2}$ and $2s_{1/2}$ neutron subshells.
 \hbox{Table \ref{tab:sn_rates}} shows the total OMC rates
 calculated with two sets of single-particle
 potential parameters \cite{Pon79}. It can be seen that the dependence of the
 total capture rates on the Woods-Saxon potential parameters is stronger 
 than its dependence on the residual interaction \hbox{coupling} 
 constants. The calculated rates show also a strong dependence on the neutron 
 excess of the target nucleus. 

\begin{sloppypar}
 The experimental value of the capture rate measured for natural $\nucl{}{Sn}$ 
 is \hbox{$106.7 \cdot 10^{5} \, {\rm s}^{-1} $} \cite{SMR1987}. Considering 
 that $\nucl{118,120}{Sn}$ contributes more than 50\% to the natural isotope 
 mixture, shows that the Woods-Saxon model parameters fitted to $A=121, Z=51$ 
 give a better description of the total rates. We can say that the agreement
 between the theoretical capture rates and the experiment is quite good.
\end{sloppypar}

\subsection{Heavy Nuclei with Large Neutron Excess. 
 $\nucl{\bf 140}{\bf Ce}$ and $\nucl{\bf 208}{\bf Pb}$}

 The most important observation to be made in the case of heavier nuclei is 
 (see \hbox{Fig.\ \ref{fig:trates-fig2}}) the large theoretical overestimation 
 of the rates. The difference between the calculated total 
 OMC rates becomes larger with increasing mass number. 
 \hbox{Table \ref{tab:ce140rates}} shows the results for $\nucl{140}{Ce}$ and 
 \hbox{Table \ref{tab:pb208rates}} those for $\nucl{208}{Pb}$. The rates 
 calculated with the residual interaction ($\ref{eq:rL}$) are smaller than 
 those obtained with the interaction ($\ref{eq:dudr}$). This difference comes 
 about mainly due to capture populating the high-excited $1^{+}$ states 
 which is absent in calculations using the interaction ($\ref{eq:rL}$). The 
 experimental energies of the collective IAS and GT ($\sigma t^{-}$) states 
 can be reproduced with the interaction ($\ref{eq:rL}$) using 
 $\kappa^{0}_{1} =-28.0/A$ to obtain $18.94$ MeV for the IAS state and using 
 $\kappa^{01}_{1} =-23.0/A$ to obtain the GT state at an energy of $19.71$ MeV.
 The corresponding experimental energies are $18.8$ MeV for the IAS state and 
 $19.2$ MeV for the GT state respectively. More than $80\%$ of the total 
 GT strength sits at the peak of the distribution. Using the interaction 
 ($\ref{eq:dudr}$) one obtains the collective $1^{+}$ state at $16.85$ MeV 
 using $\kappa^{01}_{1} =-0.43/A$. Approximately $50\%$ of the total 
 GT strength goes to this state and more than $30\%$ of the strength is 
 shifted to the higher $1^{+}$ states. These $1^{+}$ states with high 
 excitation energies are responsible for the fact that the collective $1^{+}$ 
 state remains in a region below $18$ MeV even if $|\kappa^{01}_{1}|$ is 
 doubled. 

 The experimental value of $\Lambda_{tot}$ for $\nucl{208}{Pb}$ is 
 $135 \cdot 10^{5} \, \hbox{\rm s}^{-1}$ \cite{SMR1987}. Thus both calculations 
 overestimate the total rate considerably. It is therefore interesting to 
 compare our results with those from previous work achieving a good agreement 
 with experiment \cite{Bun1966,UV1992}. There is, however, a difficulty in 
 performing this comparison because the authors of \cite{Bun1966,AueKl,UV1992} 
 presented their results as relative contributions to $\Lambda_{tot}$, 
 related to a specific angular momentum transfer $L$. To compare these data 
 to our results, we have to assume that the transitions to the $0^{+}$ and 
 $1^{+}$ final states proceed via the $L = 0$ transition. The transitions to 
 the $0^{-}, 1^{-}, 2^{-}$ states are accompanied by a $L = 1$ orbital 
 angular momentum, etc. With this assumption we implicitely assume that 
   $\bigl \vert [1\,J-1\,J] \bigr \vert >>
   \bigl \vert [1 \, J+1 \, J] \bigr \vert$. 
 A test shows that deleting of $ [1 \, J+1 \, J] $ reduces the corresponding 
 $\Lambda_{fi}$ by less than $10\%$. 
 The fractional contributions obtained by this procedure are given 
 in \hbox{Table \ref{tab:contrib}}. A comparison with the results of 
 \cite{Bun1966,AueKl,UV1992} shows that the main difference to our results 
 comes from the contributions of the $0^{+} \rightarrow 1^{+}$ transition. 
 This prominent role of the $0^{+} \rightarrow 1^{+}$ transition was also 
 found in a recent calculation \cite{KLV2000}. The total OMC rate for 
 $\nucl{208}{Pb}$ obtained in the present work with the interaction 
 ($\ref{eq:rL}$) 
 compares well with the value $\Lambda_{tot}(\nucl{208}{Pb}) = 161 \cdot 10^5  
 \hbox{\rm s}^{-1}$ obtained in \cite{KLV2000} using a $\delta$-function 
 residual interaction.

\section{Discussion and Conclusion}
 In this work a theoretical evaluation of the total OMC rates for medium - 
 weight and heavy spherical nuclei using QRPA was presented. For the first
 time an attempt was made to include the velocity-dependent terms, 
 evaluated with single particle wave functions having the correct asymptotic 
 behaviour. It was shown that the contribution of these terms to 
 ${\Lambda}^{\rm theor}_{\rm tot}$ is rather small. To avoid confusion, 
 some  remarks about the meaning of "velocity-dependent" terms have to be made. 
 Usually  all terms having its origin in the small components of the nucleon 
 $4$-spinors are called "velocity-dependent" terms. However, in the 
 derivation of the effective muon capture Hamiltonian \cite{BKE}, part of 
 these terms experience a transformation due to momentum conservation
\begin{equation}
  \vec{p} + \vec{\mu} = \vec{n} + \vec{\nu} ,
  \quad
  \vert \vec{\mu} \vert \approx 0 ,
  \quad \Rightarrow  \quad
  \vec{n} = \vec{p} - \vec{\nu} .
\end{equation}
 As a result, only the gradient acting on the proton wave functions is left.
 We have explicitly shown that the matrix elements with the proton
 gradient give a minor contribution to the capture rate.
 So one can conclude that the main effect of the velocity-dependent
 terms is already accounted for in the effective Hamiltonian due to 
 momentum conservation.

 Our calculations show that the total OMC rates are not very sensitive
 to the constants of the nuclear residual interactions.
 On the other hand they may strongly depend on the shape of the 
 residual interaction used in the calculations.
 The main influence on $\Lambda_{\rm tot}$ calculated with
 different residual interactions comes from the difference in
 the description of GT transitions,
 $0^{+} \rightarrow 1^{+}$, especially at higher excitation
 energies.
 
\begin{figure}
\begin{center}
\includegraphics[width=0.85\linewidth]{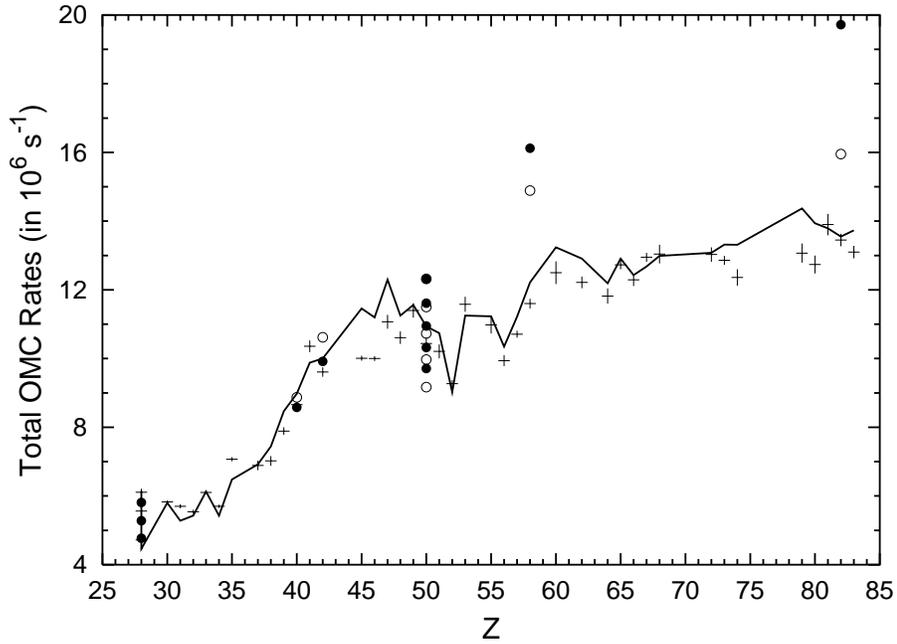}
\end{center}
\caption{Comparison between experimental and calculated total capture rates. 
 Crosses: experimental data taken from \protect{\cite{EKT1998}} and 
 \protect{\cite{SMR1987}} 
 (the vertical extension indicates the experimental error), 
 full circles: calculation with the interaction (\protect{\ref{eq:dudr}}), 
 open circles: calculation with the interaction (\protect{\ref{eq:rL}}).
 The full curve connects the rates calculated with the Goulard - Primakoff 
 formula \protect{\cite{GP1974}} with its parameters determined from 
 \protect{\cite{SMR1987}}.}
\label{fig:trates-fig2}
\end{figure}

\begin{figure}
\begin{center}
\includegraphics[width=0.85\linewidth]{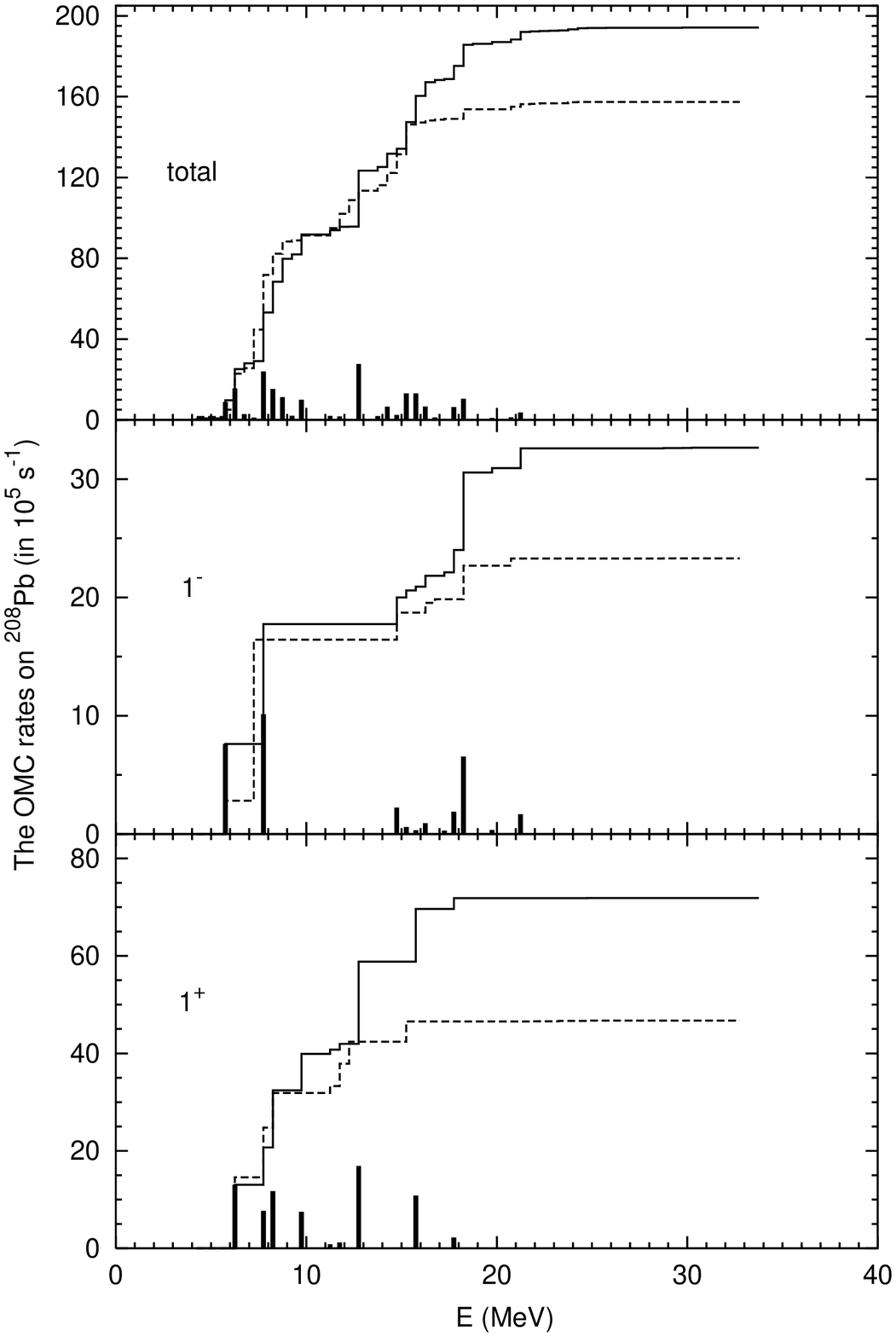}
\end{center}
\caption{%
$\nucl{208}{Pb}$ : Total and partial 
integrated capture rates up to the excitation energy E of the final nucleus. 
The partial rates are shown for the final states with $J^{\pi} = 1^{\pm}$.
Solid lines: results of the calculation with the residual 
interaction (\protect{\ref{eq:dudr}}),
dashed lines: the same with the interaction (\protect{\ref{eq:rL}}), the solid 
vertical bars show the distribution of the calulated partial rates  
over the excited states with $J^{\pi} = 1^{\pm}$. In the upper part of the 
figure the distribution over all excited states is shown.}
\label{fig:pb208omc}
\end{figure}

 A comparison of our theoretical total OMC rates with experiments shows 
 (\hbox{Table \ref{tab:total_rates}} and \hbox{Fig.\ \ref{fig:trates-fig2}}) 
 the following situation. For the medium weight nuclei ($\nucl{90}{Zr}$, 
 $\nucl{116-124}{Sn}$) a reasonable agreement between theory and experiment 
 can be achieved using the free values of $g_A$ and $g_P$. No renormalization 
 of $g_A$ is needed in this mass region.

 The ${\Lambda}^{\rm theor}_{\rm tot}$ exceed, however, considerably the 
 experimental values for the heavier nuclei $\nucl{140}{Ce}$ and 
 $\nucl{208}{Pb}$. Therefore, in order to reproduce the experimental values, 
 a renormalization of $g_A$ seems to be necessary for heavy nuclei. This 
 renormalization is model dependent; it depends in our case on the coupling
 constants and the shape of the residual interaction, as can be seen from
 Table \ref{tab:ce140rates} and \ref{tab:pb208rates}. From 
 \hbox{Fig.\ \ref{fig:trates-fig2}} one is tempted to deduce some systematic 
 deviation of ${\Lambda}^{\rm theor}_{\rm tot}$ from the experimental data. 
 This makes it impossible to draw any definite conclusion on the necessity 
 of a quenching of $g_A$ for heavier nuclei. It seems that the nuclear model 
 used in this investigation reaches its limits of useful application so that 
 further theoretical studies are necessary. The 
 widespread belief \cite{UV1992} that a theoretical description of total 
 OMC rates faces no particular problem seems to be untenable.

\newpage
\section*{Acknowledgements}

We would like to thank R. Rosenfelder for many helpful discussions and 
improvements while reading the text. Two of us, V.~A.\ Kuzmin and 
T.~V. Tetereva, would like to express their gratitude for the support obtained 
from PSI which made several visits to this institute possible.

\clearpage

\begin{landscape}
\begin{table}
\caption{Properties of charge-exchange $1^{+}$ excitations in
 $\nucl{90}{Zr}$ calculated with two variants of residual
 interactions.}
\label{tab:zr90gt}
$$
\begin{tabular}{| c | c | c | c | c | c | c | c | c |}
\hline
 {} &
 \multicolumn{4}{| c |}{ $ \sigma t^{-} $ as in $(p,n)$ reaction } &
 \multicolumn{4}{| c |}{ $ \sigma t^{+} $ as in $(n,p)$ reaction } \\
\cline{2-9}
 $\kappa^{01}_{1} \! A$ &
   Energy of &
 \multicolumn{3}{| c |}{ $B^{-}(GT)$ } &
   Energy of &
 \multicolumn{3}{| c |}{ $B^{+}(GT)$ } \\
 \cline{3-5}
 \cline{7-9}
 {} &
 maximum &
 total &
 in max.\ &
 below max.\ &
 maximum &
 total &
 in max.\ &
 below max.\ \\
\hline
  0.00   &
 11.84   &
 32.06   &
 16.96   &
 13.79   &
  5.34   &
  3.32   &
  2.21   &
  0.00   \\
\multicolumn{9}{| c |}{ $f(r) = dU/dr$ } \\
 -0.23   &
 14.93   &
 31.89   &
 20.88   &
  6.23   &
  5.62   &
  3.15   &
  1.31   &
  0.00   \\
 -0.33   &
 15.76   &
 32.03   &
 19.88   &
  4.67   &
  5.68   &
  3.30   &
  1.14   &
  0.00   \\
 -0.43   &
 16.36   &
 32.21   &
 18.41   &
  3.74   &
  5.73   &
  3.47   &
  1.03   &
  0.00   \\
\multicolumn{9}{| c |}{ $f_{L}(r) = r^{L}$ } \\
 -23.0   &
 15.73   &
 30.63   &
 23.83   &
  5.32   &
  5.63   &
  1.89   &
  1.12   &
  0.00   \\
 -25.0   &
 16.09   &
 30.56   &
 23.98   &
  4.86   &
  5.65   &
  1.82   &
  1.07   &
  0.00   \\
 -28.0   &
 16.63   &
 30.05   &
 23.86   &
  4.28   &
  5.67   &
  1.72   &
  0.99   &
  0.00   \\
\hline
\end{tabular}
$$
\end{table}

\begin{table}
\caption{Properties of charge-exchange $0^{+}$ excitations in
 $\nucl{90}{Zr}$ calculated with two variants of residual
 interactions.}
\label{tab:zr90ia}
$$
\begin{tabular}{| c | c | c | c | c | c | c | c | c |}
\hline
 {} &
 \multicolumn{4}{| c |}{ $ t^{-} $  } &
 \multicolumn{4}{| c |}{ $ t^{+} $  } \\
\cline{2-9}
 $\kappa^{0}_{1} \! A$ &
  Energy of &
 \multicolumn{3}{| c |}{ $B^{-}$ } &
  Energy of &
 \multicolumn{3}{| c |}{ $B^{+}$ } \\
 \cline{3-5}
 \cline{7-9}
 {} &
 maximum &
 total &
 in max.\ &
 below max.\ &
 maximum &
 total &
 in max.\ &
 below max.\ \\
\hline
   0.00  &    4.62 &  10.14  &  8.62  &  0.00  &  16.17  &  0.29  &  0.07  &  0.13  \\
\multicolumn{9}{| c |}{ $ f(r) = dU /dr $ } \\
  -0.33  &   10.99 &  10.58  &  8.30  &  0.37  &  18.08  &  0.72  &  0.55  &  0.16  \\
  -0.43  &   12.00 &  10.70  &  7.94  &  0.14  &  18.34  &  0.84  &  0.67  &  0.16  \\
  -0.53  &   12.81 &  10.81  &  7.44  &  0.07  &  18.56  &  0.96  &  0.79  &  0.15  \\
\multicolumn{9}{| c |}{ $f_{L}(r) = r^{L}$ } \\
 -23.00  &   10.35 &  10.04  &  7.87  &  1.78  &  16.91  &  0.18  &  0.05  &  0.07  \\
         &         &         &        &        &  16.20  &        &  0.04  &        \\
 -25.00  &   10.71 &  10.03  &  8.29  &  1.32  &  16.92  &  0.18  &  0.05  &  0.07  \\
         &         &         &        &        &  16.20  &        &  0.04  &        \\
 -28.00  &   11.28 &  10.02  &  8.68  &  0.88  &  16.92  &  0.17  &  0.05  &  0.07  \\
         &         &         &        &        &  16.20  &        &  0.04  &        \\
 -31.00  &   11.87 &  10.02  &  8.88  &  0.62  &  16.92  &  0.16  &  0.05  &  0.06  \\
         &         &         &        &        &  16.20  &        &  0.04  &        \\
 -34.00  &   12.46 &  10.01  &  8.96  &  0.45  &  16.93  &  0.15  &  0.05  &  0.06  \\
         &         &         &        &        &  16.20  &        &  0.04  &        \\
 -37.00  &   13.06 &  10.00  &  8.97  &  0.34  &  16.93  &  0.15  &  0.04  &  0.05  \\
         &         &         &        &        &  16.21  &        &  0.03  &        \\
\hline
\end{tabular}
$$
\end{table}
\end{landscape}

\begin{table}
\caption{%
 The rates of OMC (in $10^{5} \, {\rm s}^{-1}$)
 on $\nucl{90}{Zr}$ summed over the final states with
 specific spin and parity $J^{\pi}$. The second line gives the 
 contribution of each final state in \%.}
\label{tab:zr90rates}
\bigskip\begin{center}
\begin{tabular}{ | c | r | r | r | r | r | r | r | r | r | }
\hline
 $\kappa^{LJ}_{1} \! A $   &
 \multicolumn{8}{| c |}{ final states, $J^{\pi}$ } & total  \\
\cline{2-9}
   {}      &
 $ 0^{+} $ &
 $ 0^{-} $ &
 $ 1^{+} $ &
 $ 1^{-} $ &
 $ 2^{+} $ &
 $ 2^{-} $ &
 $ 3^{+} $ &
 $ 3^{-} $ &
  rate     \\
\hline
   0.00  &
    4.9  &   
    3.6  &   
   24.3  &   
   43.4  &   
   15.2  &   
   19.3  &   
   11.6  &   
    3.3  &   
  125.6  \\
    {}   &
    3.9  &   
    2.8  &   
   19.3  &   
   34.6  &   
   12.1  &   
   15.4  &   
    9.2  &   
    2.6  &   
  100.0\%  \\
\multicolumn{10}{| c |}{ $f(r) = dU / dr $ } \\
%
  -0.23  &
    5.3  &   
    2.2  &   
   28.3  &   
   27.3  &   
    8.5  &   
   12.3  &   
    5.5  &   
    1.4  &   
   90.8  \\
    {}   &
    5.9  &   
    2.5  &   
   31.2  &   
   30.0  &   
    9.3  &   
   13.5  &   
    6.0  &   
    1.6  &   
  100.0\%  \\
%
  -0.33  &
    5.3  &   
    2.2  &   
   29.1  &   
   25.1  &   
    7.4  &   
   11.2  &   
    4.4  &   
    1.2  &   
   85.8  \\
    {}   &
    6.2  &   
    2.5  &   
   33.9  &   
   29.2  &   
    8.6  &   
   13.1  &   
    5.2  &   
    1.4  &   
  100.0\%  \\
%
  -0.43  &
    5.3  &   
    2.2  &   
   29.9  &   
   23.5  &   
    6.6  &   
   10.5  &   
    3.7  &   
    1.0  &   
   82.8  \\
    {}   &
    6.4  &   
    2.6  &   
   36.1  &   
   28.4  &   
    8.0  &   
   12.7  &   
    4.5  &   
    1.2  &   
  100.0\%  \\
\multicolumn{10}{| c |}{ $f_{L}(r) = r^{L}$ } \\
%
  -23.0  &
    4.7  &   
    1.9  &   
   23.4  &   
   27.2  &   
   10.2  &   
   12.2  &   
    7.1  &   
    1.9  &   
   88.7   \\
    {}   &
    5.3  &   
    2.2  &   
   26.4  &   
   30.6  &   
   11.5  &   
   13.8  &   
    8.0  &   
    2.2  &   
  100.0\% \\
%
  -25.0  &
    4.7  &   
    1.9  &   
   23.2  &   
   27.2  &   
   10.0  &   
   11.8  &   
    6.9  &   
    1.9  &   
   86.9  \\
    {}   &
    5.5  &   
    2.2  &   
   26.7  &   
   26.5  &   
   11.5  &   
   13.6  &   
    7.9  &   
    2.2  &   
  100.0\%  \\
%
  -28.0  &
    4.7  &   
    1.8  &   
   23.0  &   
   25.7  &   
    9.6  &   
   11.3  &   
    6.5  &   
    1.8  &   
   84.4  \\
    {}   &
    5.6  &   
    2.1  &   
   27.2  &   
   30.4  &   
   11.4  &   
   13.4  &   
    7.8  &   
    2.1  &   
  100.0\%  \\
\hline
\end{tabular}
\end{center}
\end{table}
%

\begin{landscape}
\begin{table}
\caption{Properties of charge-exchange $1^{+}$ excitations in
  $\nucl{92}{Mo}$ calculated with two different nuclear residual
  interactions.}
\label{tab:mo92gt}
\bigskip\begin{center}
\begin{tabular}{| r | r | r | r | r | r | r | r | r |}
\hline
 {} &
 \multicolumn{4}{| c |}{ $ \sigma t^{-} $ as in $(p,n)$ reaction } &
 \multicolumn{4}{| c |}{ $ \sigma t^{+} $ as in $(n,p)$ reaction } \\
\cline{2-9}
 $\kappa^{01}_{1} \! A$ &
   Energy of &
 \multicolumn{3}{| c |}{ $B^{-}(GT)$ } &
   Energy of &
 \multicolumn{3}{| c |}{ $B^{+}(GT)$ } \\
 \cline{3-5}
 \cline{7-9}
 {} &
 maximum &
 total &
 in max.\ &
 below max.\ &
 maximum &
 total &
 in max.\ &
 below max.\ \\
\hline
  0.00  &
 11.85  &
 28.79  &
 16.94  &
 10.70  &
  4.44  &
  5.96  &
  4.75  &
  0.00  \\
\multicolumn{9}{| c |}{ $ f(r) = dU / dr $ } \\
 -0.23  &
 14.98  &
 27.42  &
 11.61  &
 12.17  &
  5.04  &
  4.59  &
  2.92  &
  0.00  \\
 -0.33  &
 15.61  &
 27.28  &
 16.94  &
  4.83  &
  5.19  &
  4.45  &
  2.57  &
  0.00  \\
 -0.43  &
 16.15  &
 27.23  &
 16.50  &
  3.42  &
  5.29  &
  4.40  &
  2.33  &
  0.00  \\
\multicolumn{9}{| c |}{ $f_{L}(r) = r^{L}$ } \\
 -23.0  &
 15.54  &
 26.36  &
 18.36  &
  6.86  &
  5.07  &
  3.53  &
  2.56  &
  0.00  \\
 -25.0  &
 15.81  &
 26.23  &
 19.40  &
  5.61  &
  5.11  &
  3.40  &
  2.45  &
  0.00  \\
 -28.0  &
 16.25  &
 26.06  &
 20.20  &
  4.47  &
  5.16  &
  2.23  &
  2.29  &
  0.00  \\
\hline
\end{tabular}
\end{center}
\end{table}
\end{landscape}
\begin{table}
\caption{%
 The rates of OMC (in $10^{5} \, {\rm s}^{-1}$)
 on $\nucl{92}{Mo}$ summed over the final states with
 specific spin and parity $J^{\pi}$. The second line gives the 
 contribution of each final state in \%.}
\label{tab:mo92rates}
\bigskip\begin{center}
\begin{tabular}{ |  c | r | r | r | r | r | r | r | r | r | }
\hline
 $\kappa^{LJ}_{1} \! A $   &
 \multicolumn{8}{| c |}{ final state $J^{\pi}$ } & total  \\
\cline{2-9}
    {}     &
 $ 0^{+} $ &
 $ 0^{-} $ &
 $ 1^{+} $ &
 $ 1^{-} $ &
 $ 2^{+} $ &
 $ 2^{-} $ &
 $ 3^{+} $ &
 $ 3^{-} $ &
  rate     \\
\hline
   0.00  &
    5.3  &   
    4.1  &   
   27.4  &   
   52.2  &   
   18.1  &   
   24.2  &   
   13.5  &   
    3.8  &   
  148.7  \\
    {}   &
    3.6  &   
    2.8  &   
   18.4  &   
   35.1  &   
   12.2  &   
   16.3  &   
    9.1  &   
    2.6  &   
  100.0\%  \\
\multicolumn{10}{| c |}{ $f(r) = dU / dr $ } \\
%
  -0.23  &
    5.6  &   
    2.5  &   
   30.5  &   
   33.3  &   
   10.1  &   
   15.6  &   
    6.5  &   
    1.7  &   
  105.8  \\
    {}   &
    5.3  &   
    2.4  &   
   28.8  &   
   31.5  &   
    9.5  &   
   14.7  &   
    6.2  &   
    1.6  &   
  100.0\%  \\
%
  -0.33  &
    5.6  &   
    2.4  &   
   31.0  &   
   30.6  &   
    8.8  &   
   14.1  &   
    5.3  &   
    1.4  &   
   99.2  \\
    {}   &
    5.6  &   
    2.4  &   
   31.3  &   
   30.8  &   
    8.8  &   
   14.2  &   
    5.4  &   
    1.4  &   
  100.0\%  \\
%
  -0.43  &
    5.6  &   
    2.4  &   
   31.5  &   
   28.7  &   
    7.8  &   
   13.2  &   
    4.5  &   
    1.2  &   
   95.0  \\
    {}   &
    5.9  &   
    2.6  &   
   33.2  &   
   30.3  &   
    8.3  &   
   13.9  &   
    4.8  &   
    1.8  &   
  100.0\%  \\
\multicolumn{10}{| c |}{ $f_{L}(r) = r^{L}$ } \\
%
  -23.0  &
    5.1  &   
    2.3  &   
   26.2  &   
   33.6  &   
   12.3  &   
   15.9  &   
    8.5  &   
    2.3  &   
  106.2   \\
    {}   &
    4.8  &   
    2.1  &   
   24.6  &   
   31.6  &   
   11.6  &   
   15.0  &   
    8.0  &   
    2.1  &   
  100.0\% \\
%
  -25.0  &
    5.1  &   
    2.2  &   
   26.0  &   
   32.9  &   
   12.0  &   
   15.5  &   
    8.2  &   
    2.2  &   
  104.1  \\
    {}   &
    4.9  &   
    2.1  &   
   24.9  &   
   31.6  &   
   11.5  &   
   14.9  &   
    7.9  &   
    2.2  &   
  100.0\%  \\
%
  -28.0  &
    5.1  &   
    2.1  &   
   25.7  &   
   31.8  &   
   11.6  &   
   14.9  &   
    7.9  &   
    2.2  &   
  101.2  \\
    {}   &
    5.1  &   
    2.1  &   
   25.4  &   
   31.5  &   
   11.5  &   
   14.7  &   
    7.8  &   
    2.1  &   
  100.0\%  \\
\hline
\end{tabular}
\end{center}
\end{table}
\clearpage

\begin{table}
\caption{Total OMC capture rates on Sn isotopes
 (in $10^{5} \, \hbox{\rm s}^{-1}$).}
\label{tab:sn_rates}
\bigskip\begin{center}
\begin{tabular}{ | c | r | r | r | r | r | r | }
\hline
 Target & $\kappa^{01}_{1} \, A $ &
  \multicolumn{2}{| c |}{ SW parameters } &
   $\kappa^{01}_{1} \, A $ &
  \multicolumn{2}{| c |}{ SW parameters } \\
 \cline{3-4}
 \cline{6-7}
 nucleus & for (\ref{eq:dudr}) & 115,49 & 121,51 &
           for (\ref{eq:rL}) & 115,49 & 121,51 \\
\hline
 $\nucl{116}{Sn}$ &  -0.23  &  139.2  &  130.1  &  -23.0  &  141.7  &  123.0  \\
    {}            &  -0.33  &  130.5  &  123.1  &  -25.0  &  138.9  &  120.8  \\
    {}            &  -0.43  &  124.9  &  119.0  &  -28.0  &  135.2  &  117.4  \\
\hline
 $\nucl{118}{Sn}$ &  -0.23  &  130.0  &  122.1  &  -23.0  &  131.9  &  115.0  \\
    {}            &  -0.33  &  122.1  &  116.1  &  -25.0  &  129.4  &  112.7  \\
    {}            &  -0.43  &  117.1  &  112.4  &  -28.0  &  125.9  &  109.5  \\
\hline
 $\nucl{120}{Sn}$ &  -0.23  &  121.2  &  111.8  &  -23.0  &  122.6  &  107.3  \\
    {}            &  -0.33  &  114.2  &  109.5  &  -25.0  &  120.3  &  104.1  \\
    {}            &  -0.43  &  109.8  &  106.3  &  -28.0  &  117.2  &  102.1  \\
\hline
 $\nucl{122}{Sn}$ &  -0.23  &  113.2  &  107.8  &  -23.0  &  114.0  &   99.7  \\
    {}            &  -0.33  &  106.9  &  103.2  &  -25.0  &  118.0  &   97.7  \\
    {}            &  -0.43  &  103.0  &  100.6  &  -28.0  &  108.0  &   95.0  \\
\hline
 $\nucl{124}{Sn}$ &  -0.23  &  105.5  &  101.3  &  -23.0  &  105.5  &   91.7  \\
    {}            &  -0.33  &   99.9  &   97.1  &  -25.0  &  103.4  &   89.9  \\
    {}            &  -0.43  &   96.5  &   95.1  &  -28.0  &  100.7  &   88.3  \\
\hline
\end{tabular}
\end{center}
\end{table}

\begin{table}
\caption{%
 The rates of OMC (in $10^{5} \, {\rm s}^{-1}$)
 on $\nucl{140}{Ce}$ summed over final states with
 specific spin and parity $J^{\pi}$. The second line gives the 
 contribution of each final state in \%.}
\label{tab:ce140rates}
\bigskip\begin{center}
\begin{tabular}{ |  c | r | r | r | r | r | r | r | r | r | }
\hline
 $\kappa^{LJ}_{1} \! A $   &
 \multicolumn{8}{| c |}{ final state $J^{\pi}$ } & total  \\
\cline{2-9}
     {}    &
 $ 0^{+} $ &
 $ 0^{-} $ &
 $ 1^{+} $ &
 $ 1^{-} $ &
 $ 2^{+} $ &
 $ 2^{-} $ &
 $ 3^{+} $ &
 $ 3^{-} $ &
  rate     \\
\hline
\multicolumn{10}{| c |}{ $f(r) = dU /dr $ } \\
%
   -0.23  &
    11.7  &  
     3.8  &  
    55.4  &  
    39.9  &  
    20.2  &  
    21.0  &  
    10.3  &  
     3.5  &  
   165.7  \\
    {} &
     7.0  &  
     2.3  &  
    33.4  &  
    24.1  &  
    12.2  &  
    12.7  &  
     6.2  &  
     2.1  &  
  100.0\%  \\
   -0.33  &
    11.7  &  
     3.5  &  
    57.4  &  
    38.0  &  
    17.9  &  
    21.3  &  
     8.7  &  
     2.9  &  
   161.2  \\
    {} &
     7.2  &  
     2.2  &  
    35.6  &  
    23.5  &  
    11.1  &  
    13.2  &  
     5.4  &  
     1.8  &  
   100.0\%  \\
   -0.43  &
    11.7  &  
     3.4  &  
    59.2  &  
    36.7  &  
    16.2  &  
    21.7  &  
     7.7  &  
     2.5  &  
   159.0  \\
    {} &
     7.3  &  
     2.1  &  
    37.3  &  
    23.1  &  
    10.2  &  
    13.7  &  
     4.8  &  
     1.6  &  
   100.0\%  \\
\multicolumn{10}{| c |}{ $f_{L}(r) = r^{L}$ } \\
%
   -23.0  &
     9.8  &  
     3.4  &  
    41.4  &  
    35.8  &  
    25.2  &  
    15.1  &  
    12.9  &  
     5.1  &  
   148.9  \\
    {} &
     6.6  &  
     2.3  &  
    27.8  &  
    24.1  &  
    16.9  &  
    10.1  &  
     8.7  &  
     3.4  &  
   100.0\%  \\
   -25.0  &
     9.8  &  
     3.3  &  
    41.2  &  
    35.1  &  
    24.7  &  
    14.8  &  
    12.5  &  
     5.0  &  
   146.3  \\
    {} &
     6.7  &  
     2.2  &  
    28.1  &  
    24.0  &  
    16.9  &  
    10.1  &  
     8.6  &  
     3.4  &  
   100.0\%  \\
   -28.0  &
     9.8  &  
     3.1  &  
    40.8  &  
    34.1  &  
    23.9  &  
    14.4  &  
    12.0  &  
     4.8  &  
   142.8  \\
    {} &
     6.9  &  
     2.2  &  
    28.5  &  
    23.9  &  
    16.7  &  
    10.1  &  
     8.4  &  
     3.4  &  
   100.0\%  \\
\hline
\end{tabular}
\end{center}
\end{table}

\begin{table}
\caption{%
 The rates of OMC (in $10^{5} \, {\rm s}^{-1}$)
 on $\nucl{208}{Pb}$ summed over final states with
 specific spin and parity $J^{\pi}$. The second line gives the 
 contribution of each final state in \%.}
\label{tab:pb208rates}
\bigskip\begin{center}
\begin{tabular}{ |  c | r | r | r | r | r | r | r | r | r | }
\hline
 $\kappa^{LJ}_{1} \! A $   &
 \multicolumn{8}{| c |}{ final state $J^{\pi}$ } & total  \\
\cline{2-9}
    {}     &
 $ 0^{+} $ &
 $ 0^{-} $ &
 $ 1^{+} $ &
 $ 1^{-} $ &
 $ 2^{+} $ &
 $ 2^{-} $ &
 $ 3^{+} $ &
 $ 3^{-} $ &
  rate     \\
\hline
\multicolumn{10}{| c |}{ $ f(r) = dU / dr $ } \\
%
   -0.23  &
    15.7  &  
     3.1  &  
    70.5  &  
    33.8  &  
    29.7  &  
    29.5  &  
    10.4  &  
     5.8  &  
   198.6  \\
    {} &
     7.9  &  
     1.6  &  
    35.5  &  
    17.0  &  
    15.0  &  
    14.9  &  
     5.2  &  
     2.9  &  
  100.0\%  \\
   -0.33  &
    15.7  &  
     2.7  &  
    72.5  &  
    33.1  &  
    26.9  &  
    31.8  &  
     9.5  &  
     4.9  &  
   197.2  \\
    {} &
     8.0  &  
     1.4  &  
    36.8  &  
    16.8  &  
    13.7  &  
    16.1  &  
     4.8  &  
     2.5  &  
   100.0\%  \\
   -0.43  &
    15.7  &  
     2.4  &  
    71.9  &  
    32.6  &  
    25.0  &  
    33.2  &  
     9.0  &  
     4.3  &  
   194.2  \\
    {} &
     8.1  &  
     1.2  &  
    37.0  &  
    16.8  &  
    12.9  &  
    17.1  &  
     4.6  &  
     2.2  &  
   100.0\%  \\
\multicolumn{10}{| c |}{ $f_{L}(r) = r^{L}$ } \\
%
   -23.0  &
    13.5  &  
     2.8  &  
    46.9  &  
    23.8  &  
    33.7  &  
    19.9  &  
    11.0  &  
     8.0  &  
   159.5  \\
    {} &
     8.5  &  
     1.7  &  
    29.4  &  
    14.9  &  
    21.1  &  
    12.4  &  
     6.9  &  
     5.0  &  
   100.0\%  \\
   -25.0  &
    13.5  &  
     2.6  &  
    46.7  &  
    23.3  &  
    33.0  &  
    19.8  &  
    10.7  &  
     7.8  &  
   157.4  \\
    {} &
     8.6  &  
     1.7  &  
    29.7  &  
    14.8  &  
    21.0  &  
    12.6  &  
     6.8  &  
     5.0  &  
   100.0\%  \\
   -28.0  &
    13.5  &  
     2.4  &  
    46.5  &  
    22.6  &  
    32.0  &  
    19.6  &  
    10.3  &  
     7.5  &  
   154.5  \\
    {} &
     8.7  &  
     1.6  &  
    30.1  &  
    14.7  &  
    20.7  &  
    12.7  &  
     6.7  &  
     4.9  &  
   100.0\%  \\
\hline
\end{tabular}
\end{center}
\end{table}

\newpage
\begin{table}
\caption{%
 Fractional contributions of different multipoles
 to $\Lambda_{\rm tot}$ for $\nucl{208}{Pb}$
 (either contributions of the orbital momentum transfer
 or contributions of the transitions to the states with
 specific $J^{\pi}$).
 }
\label{tab:contrib}
\bigskip\begin{center}
\begin{tabular}{ | c | c | c | c | c | c | c | c | c | }
\hline
 Reference &
 \multicolumn{8}{| c |}{ $\Delta L$ } \\
 \cline{2-9}
 {} &
 \multicolumn{2}{| c |}{ 0 } &
 \multicolumn{3}{| c |}{ 1 } &
 \multicolumn{2}{| c |}{ 2 } &
                         3   \\
 \hline
 \cite{Bun1966} &
 \multicolumn{2}{| c |}{ 26 } &
 \multicolumn{3}{| c |}{ 14 } &
 \multicolumn{2}{| c |}{ 48 } &
                         12   \\
 \cite{AueKl}   &
 \multicolumn{2}{| c |}{ 29 } &
 \multicolumn{3}{| c |}{ 13 } &
 \multicolumn{2}{| c |}{ 52 } &
                          7   \\
 \cite{UV1992}  &
 \multicolumn{2}{| c |}{ 23 } &
 \multicolumn{3}{| c |}{ 34 } &
 \multicolumn{2}{| c |}{ 34 } &
                          8   \\
 (a)           &
 \multicolumn{2}{| c |}{ $ \simeq 45 $ } &
 \multicolumn{3}{| c |}{ $ \simeq 35 $ } &
 \multicolumn{2}{| c |}{ $ \simeq 18 $ } &
                         $ \simeq 2  $   \\
 (b)           &
 \multicolumn{2}{| c |}{ $ \simeq 38 $ } &
 \multicolumn{3}{| c |}{ $ \simeq 29 $ } &
 \multicolumn{2}{| c |}{ $ \simeq 27 $ } &
                         $ \simeq 5  $   \\
 \hline
 {} &
 \multicolumn{8}{| c |}{  $J^{\pi}$ } \\
 \cline{2-9}
 {} &
 $ 0^{+} $ &
 $ 1^{+} $ &
 $ 0^{-} $ &
 $ 1^{-} $ &
 $ 2^{-} $ &
 $ 2^{+} $ &
 $ 3^{+} $ &
 $ 3^{-} $ \\
\hline
 \cite{KLV2000} &
 $  5  $ &  
 $ 36  $ &  
 $  6  $ &  
 $ 14  $ &  
 $ 12  $ &  
 $ 28  $ &  
 $  4  $ &  
 $  2  $ \\ 
 (a)     &
 $  8  $ &  
 $ 37  $ &  
 $  2  $ &  
 $ 17  $ &  
 $ 16  $ &  
 $ 14  $ &  
 $  5  $ &  
 $  3  $ \\ 
 (b)     &
 $  8  $ &  
 $ 30  $ &  
 $  2  $ &  
 $ 15  $ &  
 $ 12  $ &  
 $ 21  $ &  
 $  7  $ &  
 $  5  $ \\ 
\hline
\end{tabular}
\end{center}
  (a) -- present paper, calculations with the residual interaction
     (\ref{eq:dudr}), $\kappa^{LJ}_{1} \cdot A = -0.43 $; \\
  (b) -- present paper, calculations with the residual interaction
     (\ref{eq:rL}), $\kappa^{LJ}_{1} \cdot A = -25.0. $
\end{table}

%
\clearpage

\begin{table}
 \caption{ %
 Summary of results : total OMC rates (in $ 10^{5} \, {\rm s}^{-1}$)
 calculated for $\geff{P}/\geff{A} = 6.0 $ with two
 different radial form factors of the nuclear residual 
 interaction. }
\label{tab:total_rates}
\bigskip \begin{center}
\begin{tabular}{ | c | r | r | r | r | r | r | r | }
\hline
 Target &
 \multicolumn{3}{| c |}{ $ \kappa^{LJ}_{1} A $ for $ d U / d r $ } &
 \multicolumn{3}{| c |}{ $ \kappa^{LJ}_{1} A $ for $ r^{L} $  }    &
 expr. \\
\cline{2-7}
 nucleus & $-0.23$ & $-0.33$ & $-0.43$ & $-23.0$ & $-25.0$ & $-28.0$ &
 \cite{SMR1987} \quad \\
\hline
 $\nucl{90}{Zr}$ &
    90.8         &
    85.8         &
    82.8         &
    88.7         &
    86.8         &
    84.4         &
    86.6         \\
 $\nucl{92}{Mo}$ &
   105.8         &
    99.2         &
    95.0         &
   106.2         &
   104.1         &
   101.2         &
    92.2        \\
 $\nucl{116}{Sn}$ &
   130.1         &
   123.2         &
   119.0         &
   123.2         &
   120.8         &
   117.4         &
    {}         \\
 $\nucl{118}{Sn}$ &
   122.1         &
   116.1         &
   112.4         &
   115.0         &
   112.7         &
   109.5         &
    {}         \\
 $\nucl{120}{Sn}$ &
   114.8         &
   109.5         &
   106.3         &
   107.3         &
   105.1         &
   102.1         &
   106.7         \\
 $\nucl{122}{Sn}$ &
   107.8         &
   103.2         &
   100.6         &
    99.7         &
    97.7         &
    95.0         &
    {}         \\
 $\nucl{124}{Sn}$ &
   101.3         &
    97.1         &
    95.1         &
    91.7         &
    89.9         &
    88.3         &
    {}         \\
 $\nucl{140}{Ce}$ &
   165.7         &
   161.2         &
   159.0         &
   148.9         &
   146.3         &
   142.8         &
   114.4         \\
 $\nucl{208}{Pb}$ &
   198.6         &
   197.2         &
   194.2         &
   159.5         &
   157.4         &
   154.5         &
   134.5        \\
\hline
\end{tabular}
\end{center}
\end{table}

\clearpage

\newpage

\newpage
\end{document}